\documentclass[aps,pra,amsmath,twocolumn,amssymb,floatfixng,showpacs,
superscriptaddress,nofootinbib]{revtex4}
\pdfoutput=1
\usepackage[dvips]{graphics}
\usepackage{bm}
\usepackage{float}
\usepackage{epsfig}
\usepackage{enumerate}
\usepackage{amsmath}
\usepackage{color}
\usepackage{graphicx}
\usepackage[colorlinks=true,linktoc=page,linkcolor=red,citecolor=blue]{hyperref}

\newcommand\bea{\begin{eqnarray}}
\newcommand\eea{\end{eqnarray}}
\newcommand\beq{\begin{equation}}  
\newcommand\eeq{\end{equation}}

\begin{document}

\title{Leggett-Garg inequality in Markovian quantum dynamics: role of temporal sequencing of coupling to bath}

\author{Sayan Ghosh}%
 \email{sayanbghosh@gmail.com}

\author{Anant V. Varma}%
 \email{anantvijay.cct@gmail.com}
 
\author{Sourin Das}
 \email{sourin@iiserkol.ac.in, sdas.du@gmail.com}

\affiliation{%
Indian Institute of Science Education \& Research Kolkata,
Mohanpur, Nadia - 741 246, 
West Bengal, India
% IISER-Kolkata, Mohanpur,741246, West Bengal, India.
}%

\date{\today}
	
\begin{abstract}
 We study Leggett-Garg inequalities (LGIs) for a  two level system (TLS)  undergoing Markovian dynamics described by unital maps.  We find analytic expression of LG parameter $K_{3}$ (simplest variant of LGIs) in terms of the parameters of two distinct unital  maps representing time evolution for intervals: $t_{1}$ to $t_{2}$ and $t_{2}$ to $t_{3}$. We show that the maximum violation of LGI for these  maps can never exceed well known L\"{u}ders bound of $K_{3}^{L\ddot{u}ders}=3/2$ over the full parameter space.
 We further show that if the map for the time interval $t_{1}$ to $t_{2}$ is non-unitary unital then irrespective of the choice of the  map for interval $t_{2}$ to $t_{3}$ we can never reach L\"{u}ders bound.
On the other hand, if the measurement operator eigenstates remain pure upon evolution from $t_{1}$ to $t_{2}$, then depending on the degree of decoherence induced by the unital map  for the interval $t_{2}$ to $t_{3}$ we may or may not obtain L\"{u}ders bound. Specifically, we find that if the unital map for interval $t_{2}$ to $t_{3}$ leads to the shrinking of the Bloch vector beyond half of its unit length, then achieving the bound $K_{3}^{L\ddot{u}ders}$ is not possible. Hence our findings not only establish a threshold for decoherence which will allow for $K_{3} = K_{3}^{L\ddot{u}ders}$, but also demonstrate the importance of temporal sequencing of the exposure of a TLS to Markovian baths in obtaining L\"{u}ders bound.
 
 %We demonstrate the role of temporal sequencing of the two dynamical maps for this sub-set of parameter space, where one dynamical map is 'unitary like' and other map is 'non-unitary unital'. We find that 'unitary like' maps should precede 'non-unitary unital' maps in order to violate LGI upto Lüder's bound .

\end{abstract}
	
\maketitle

\section{\label{intro}Introduction}

 LGIs are the temporal analogs of Bell's inequality \citep{Bell1964, Peres1999, Mahler1993, Kim2006, Budroni2013} and provide a test for quantum mechanics at a macroscopic scale via violating the bounds on the temporal correlations \cite{LG1985,Leggett2008,Legget2002}. Violation of LGIs indicates a breakdown of either or both of (a) macroscopic realism and (b) non-invasive measurability.  These two postulates underlie the basic construction of LGIs ~\cite{LGIreview2014}, and are based on our intuition of the classical world. Thus LGI violation can be considered as an indication  of non-classical or quantum behavior under appropriate experimental circumstances~\cite{Mahler1993,Wilde2012}. Among the many different variants of these inequalities the simplest three-time measurements scenario can be written as: $-3 \leq K_{3}=C_{12}+C_{23}-C_{13}\leq 1$. The two time-correlation $ C_{ij}=\sum_{Q(t_i),Q(t_j)=\pm 1} Q(t_i) Q(t_j)P_{ij}(Q(t_i),Q(t_j))$, where $Q(t_i),Q(t_j)$ represent the outcomes of the projective measurements of the observable $Q$ at times $t_i$ and $t_j$ respectively and $P_{ij}(Q(t_i),Q(t_j))$ represents the joint probability of the outcome of quantum measurement performed at time $t_i$ and  $t_j$ to be $Q(t_i)$, $Q(t_j)$ respectively. The maximum quantum bound of $K_3$ for an $N$ level system is $3/2$ which is known L\"{u}ders bound ~\cite{Emary2014}. Violation of this bound for an $N$  level quantum system, where $N>2$ is possible provided further degeneracy breaking measurements are performed \cite{Emary2014,Pan} but the violation of L\"{u}ders bound for $N=2$ $i.e.$ a two level system (TLS) is impossible within the unitary dynamics.\\
 
Albeit in practice, due to the coupling with the environment, the dynamics of any quantum system will deviate from the ideal unitary dynamics and become non-unitary.  All valid non-unitary dynamics (that map
 a density matrix to another density matrix) can be defined using completely positive and trace preserving (CPTP) maps \cite{sudarshan, plenio1,lock,vedral,lupo,lian}. Kraus \cite{kraus1,kraus2} and Choi \cite{choi} demonstrated that there exists a set of operators $A_{k}$ for any CPTP dynamical map $\Lambda$ acting on a state $\rho$ such that $\rho(t) = \Lambda(\rho) = \sum_{k} A_{k} \  \rho \ A_{k}^{\dagger}$, where trace preservation is defined as  $\sum_{k} \ A_{k}^{\dagger} \  A_{k}  =\mathbb{I}$. In this article we discuss the sub-set of CPTP maps which map the identity operator to itself: $\Lambda(\mathbb{I}) = \mathbb{I}$, in the context of LGI. These maps are known as unital maps $i.e.$ $\sum_{k} \  A_{k} \ A_{k}^{\dagger}   =\mathbb{I}$ \cite{sudarshan3,King-2001,Keyl-2002,mario,bacon,zanardi}. Physically, the system defined by $\rho$ undergoes decoherence but not relaxation under the action of such maps \cite{Emary}. In particular, we analyze  the dynamics defined by unital CPTP maps which are infinitesimally divisible \cite{plenio, hri} and can achieve $K_{3} = K_{3}^{L\ddot{u}ders}$.
 %with the constraint of LG parameter being $K_{3}^{L\ddot{u}ders}$. 
 Since all the CPTP maps which are infinitesimally divisible are Markovian maps, the dynamics discussed here is Markovian \cite{Sudarsan-1977}. Note that Emary has discussed LG parameter $K_{3}$ for specific unital CPTP maps \cite{Emary} in the past, whereas we find analytic expression for $K_{3}$ in terms of the parameters of any  two arbitrary unital positive maps. However, it should be noted that Emary considered  all possible measurements at different times, whereas analytical results presented here assumes a fixed measurement direction at all times. Moreover, Emary focused on the sub-set of the parameters of the dynamical maps where LGI violation $i.e.$ $K_{3}>1$ is possible, on the contrary we derive the constraints on the parameters of two positive unital maps (though the focus of the paper is restricted to CPTP maps only) solely for the case where $K_{3} = K_{3}^{L\ddot{u}ders}$. We find that a TLS exposed to an environment represented by CPTP unital and infinitesimally divisible map has a degree of tolerance to decoherence beyond which the violation of LGI upto L\"{u}ders bound is impossible. Moreover, the ordering of exposure to the decoherence (the temporal sequence of these maps) between the measurements is important and plays a crucial role in obtaining $K_{3} = K_{3}^{L\ddot{u}ders}$. We consider an explicit example where one dynamical map is such that it evolves  measurement operator eigenstates to pure states and we label it  ``unitary like'' \footnote{At least evolves  measurement operator eigenstates to pure states} and other map is ``non-unitary unital'' \footnote{This map is such that at least the eigenstates of  measurement operator do not evolve to pure states}. We provide a geometric view of sensitivity of obtaining $K_{3}^{L\ddot{u}ders}$ to the temporal sequencing of these maps between the measurements on the Bloch sphere. We  find that maximum violation of LGI can only be achieved only if the map for the time interval $t_{2}-t_{1}$ is unitary like, while the map for the time interval  $t_{3}-t_{2}$ can both be non-unitary unital (with the threshold to decoherence) or unitary like.
 
 This article is organized as follows. In section \ref{CPTP maps} we review and discuss CPTP maps and their divisibility. We  also review unital maps and introduce Stokes vector representation  to write such maps. In section \ref{unital LGI} we derive algebraic expression for LG parameter $K_{3}$ in terms of the parameters of the two unital positive maps and find the sub-set where L\"{u}ders bound can be obtained. We then illustrate the role of the ordering of the two CPTP unital and infinitesimally divisible maps between the measurements to obtain the L\"{u}ders bound. Section \ref{discussion} is dedicated to conclusion.

\section{\label{CPTP maps}  CPTP unital maps} 

When the initial interactions between system (represented by an initial state $\rho_{0}$) and the environment can be neglected then the evolution of such open system is well-described by a CPTP map. Any dynamical CPTP map $\Lambda^{t_{2},t_{1}}$ for time interval $t_{2}-t_{1}$ can always be written in the Kraus representation as \cite{kraus1,kraus2}:

\begin{equation}
\rho(t) = \Lambda^{t,0}(\rho_{0}) = \sum_{k} A_{k} \  \rho_{0} \ A_{k}^{\dagger}
\label{E1}
\end{equation}

where the Kraus operators $A_{k}$ satisfy the completeness relation: $\sum_{k} \ A_{k}^{\dagger} \  A_{k}  =\mathbb{I}$ and is a condition for trace preservation. As it has been emphasized in \cite{plenio} that infinitesimal divisibility of any CPTP map $\Lambda^{t,0}$ is defined as:

\begin{equation}
\Lambda^{t,0} = \Lambda^{t,t-\epsilon}\Lambda^{t-\epsilon,t-2 \epsilon}...\Lambda^{\epsilon,0},
\label{E2}
\end{equation}

where $\epsilon$ is infinitesimal small and all  $\{ \Lambda \} $'s  are CPTP maps, guarantees that dynamical map $\Lambda^{t,0}$ is Markovian. However, if there exist a map in the set  $\{ \Lambda \} $ in eqn. (\ref{E2}) such that it is not CPTP map, the dynamics is considered to be non-Markovian. It should be noted that in general it is a difficult task to completely parametrize all possible Markovian dynamics  and therefore we restrict our analysis to a sub-set of all possible Markovian dynamics which are unital in nature. Such dynamics corresponds to physical situations, where the system undergoes decoherence without relaxation \cite{Emary}.

\subsection{\label{unital maps}  Unital Maps}

Any dynamical map $\Lambda$ is unital if it preserves the identity element \cite{King-2001,  sudarshan3,Keyl-2002} $i.e.$ $\Lambda(\mathbb{I}) = \mathbb{I}$. This implies that a maximally mixed state (which is proportional to identity in the matrix representation) remains invariant.
In terms of Kraus representation $\Lambda$ is unital if  the condition $\sum_{k} A_{k} \ A_{k}^{\dagger}= \mathbb{I}$ is satisfied. However, Kraus representation is cumbersome to work with in general. Therefore, we follow another representation of the dynamical map $\Lambda$ defined by King and Ruskai  in \cite{King-2001} for a TLS. They showed that any linear trace preserving dynamical map $\Lambda$ for a TLS can be represented by a unique 4x4 matrix operator $\mathbb{T} = \begin{pmatrix}
1 & \mathbf{0} \\ 
\mathbf{b} &  \mathbf{\Delta}
\end{pmatrix}$, which acts on the  four component Stokes vector $(1,\vec{w}_{x},\vec{w}_{y},\vec{w}_{z})$ where $\vec{w}= (\vec{w}_{x},\vec{w}_{y},\vec{w}_{z})$ defines the Bloch vector corresponding to the quantum state of TLS, $\mathbf{0}$ and $\mathbf{b}$ are row null-vector and column vector respectively. In essence, $\mathbb{T}$ is a 4x4 matrix representation of dynamical map $\Lambda$.
Since the state of a TLS can be written as $ \rho = 1/2 \ [  \ \mathbb{I}_2 + \vec w \cdot \vec \sigma \ ]$. The action of operator $\mathbb{T}$ on the state $\rho$ is then given by:

\begin{equation}
 \Lambda(\rho)= \dfrac{1}{2}  [ \ \mathbb{I}_2 + (\mathbf{b} + \mathbf{\Delta} \vec w ) \cdot \vec \sigma  \   ]
\label{E3}
\end{equation}

The dynamical map $\Lambda$ in eqn.~(\ref{E3}) is only trace preserving. This map is unital when the column vector  $\mathbf{b}$ is a null vector. The action of a typical unital map defined by $\mathbf{\Delta}$ on the Bloch sphere results in an ellipsoid with its center being at the center of the Bloch sphere. Therefore, for a TLS  trace preserving unital map $\Lambda$ is solely defined by matrix $\mathbf{\Delta}$. To ensure the complete positivity of map $\Lambda$ it is convenient to parametrize the matrix as $\mathbf{\Delta} = R_1^T D R_2$ \cite{King-2001, Emary, Ruskai-2002, Wolf-2008}, where $R_1, R_2$ are orthogonal matrices in general and $D=\text{diag} (c_1, c_2, c_3)$ is a diagonal matrix.  A necessary and sufficient condition for such maps to be completely positive is that $ c_3 + c_2 \leq 1 + c_1$  where $1 \geq c_3 \geq c_2 \geq |c_1|$ \cite{ Wolf-2008}. Finally, infinitesimal divisibility of the map $\Lambda$ is ensured by the determinant  $\text{det}(\mathbb{T}) = \text{det}(\mathbf{\Delta}) \geq 0$ \cite{Wolf-2008, Davalos-2019}. In what follows we denote the dynamical map $\Lambda= \Lambda_{\mathbf{\Delta}}$ as it is defined by the matrix $\mathbf{\Delta}$ only when it is a unital map.

\section{\label{unital LGI}  LGI for unital maps} 

We now analyze LGI for a TLS, where the evolution between the measurements is denoted by unital map $\Lambda_{\mathbf{\Delta}}$. We utilize a simplest variant of LGI, which is based on three measurements. This variant is defined as $K_{3}=C_{12}+C_{23}-C_{13}$. Without loss of generality we assume that projective measurements are performed along z- direction (in $\sigma_{z}$ basis). We start this section by revisiting the unitary dynamics and illustrate the maximum violation of $K_{3}$. We then define two unital maps $\Lambda_{\mathbf{\Delta}_{12}}$ and $\Lambda_{\mathbf{\Delta}_{23}}$ for time intervals $t_{2}-t_{1}$ and $t_{3}-t_{2}$ respectively and derive an algebraic expression for the LG parameter $K_{3}$ in terms of the parameters of both $\mathbf{\Delta}_{12}$ and $\mathbf{\Delta}_{23}$. As we have shown later that for unital maps $K_{3}$ is independent of initial state at time $t=0$, hence $K_{3}$ is a function of the parameters of only $\mathbf{\Delta}_{12}$ and $\mathbf{\Delta}_{23}$ matrices.

\subsection{Unitary Dynamics}

The generic expression of $K_{3}$ for unitary dynamics (which can involve Hamiltonian with explicit time dependence), for a TLS can be written as: 

%\bea
%   K_{3} &=& \cos 2\theta_1 + \cos 2\theta_2 - \cos 2 \theta_1 \cos 2\theta_2
%     \nn\\
%         && - \sin 2\theta_1 \sin 2\theta_2 \cos 2 \gamma
%\label{E03}
%\eea

\begin{equation}
 \begin{split}
   K_{3} &= \cos 2\theta_1 + \cos 2\theta_2 - \cos 2 \theta_1 \cos 2\theta_2  \\
         & - \sin 2\theta_1 \sin 2\theta_2 \cos 2 \gamma  
\end{split} 
\label{E03}
\end{equation}

where $\theta_{1}$, $\theta_{2}$ and $\gamma$ are parameters related to the matrix elements of the evolution operators in the measurement basis (see Appendix A) and $C_{12}= \cos 2\theta_{1}$, $C_{23}= \cos 2\theta_{2}$ and $C_{13}=  \cos 2 \theta_1 \cos 2\theta_2 + \sin 2\theta_1 \sin 2\theta_2 \cos 2 \gamma $. 
%The value of two-time correlation functions are 
%$C_{12}= \cos 2\theta_{1}$, $C_{23}= \cos 2\theta_{2}$ and $C_{13}=  \cos 2 \theta_1 \cos 2\theta_2 + \sin 2\theta_1 \sin 2\theta_2 \cos 2 \gamma $.
In quantum mechanics maximum bound on $K_{3}$ has been shown  to be equal to $3/2$ \cite{Barberi,Kofler}, and is known as L\"{u}ders bound. Upon setting $\gamma=\pi/2$ and $\theta_{1}= \theta_{2}= \pi/6$ it is clear from the above definitions that
$K_{3}= K_{3}^{L\ddot{u}ders}=3/2$ is possible only when $C_{12}= C_{23}= -C_{13}= 1/2$. In fact unitary dynamics pins the absolute value of  the correlation $|C_{ij}|=1/2$ for maximum violation irrespective of the other details. Interestingly we find that this structure  $|C_{ij}|=1/2$ is still valid for non-unitary unital dynamics as well, in order to obtain the $K_{3}^{L\ddot{u}ders}$ and is discussed later. 

\subsection{Non-Unitary Dynamics}
 In this sub-section we first derive the algebraic expression for $K_{3}$ in terms of elements of the matrices $\mathbf{\Delta}_{12}$ and $\mathbf{\Delta}_{23}$. The measurement operator is chosen to be $\sigma_{z}$ without any loss of generality. Three correlations namely $C_{12},C_{23}$ and $C_{13}$ can be written in terms of matrix elements of the matrices $\mathbf{\Delta}_{12}$ and $\mathbf{\Delta}_{23}$ as: 
 
\begin{equation}
\begin{matrix}
    C_{12}= \mathbf{\Delta}_{12}(3,3); \\
    C_{23} = \mathbf{\Delta}_{23}(3,3); \\
    C_{13}= ( \ \mathbf{\Delta}_{23} \mathbf{\Delta}_{12} \ ) (3,3); 
\end{matrix} 
\label{E05}
\end{equation} \\

where $\mathbf{\Delta}_{ij}(3,3)$ is (3,3) element of the $3\times3$ matrix $\mathbf{\Delta}_{ij}$ and $( \ \mathbf{\Delta}_{12} \mathbf{\Delta}_{23} \ )$ is the product of two matrices $\mathbf{\Delta}_{12}$ and $\mathbf{\Delta}_{23}$. This implies that the time evolution for the time interval $t_{3}-t_{1}$ is a composition of evolution for time interval $t_{2}-t_{1}$ and $t_{3}-t_{2}$ $i.e.$ $\mathbf{\Delta}_{13}= \mathbf{\Delta}_{23}\mathbf{\Delta}_{12}$. Hence we assume that the corresponding map $\Lambda_{\mathbf{\Delta}_{13}}$ is divisible at time $t_{2}$.  Important to notice that $K_{3}$ can be completely parametrized using only the $(3,3)$ elements of these two matrices and their product. %We now re-write the  
%$( \ \mathbf{\Delta}_{12} %\mathbf{\Delta}_{23} \ ) %(3,3)$ in terms of the %elements of the individual %matrices as:

% \begin{gather*}
%    ( \ \mathbf{\Delta}_{23} \mathbf{\Delta}_{12} \ ) (3,3) = \mathbf{\Delta}_{23}(3,1) \  \mathbf{\Delta}_{12}(1,3) \\
%    + \mathbf{\Delta}_{23}(3,2) \ \mathbf{\Delta}_{12}(2,3) + \mathbf{\Delta}_{23}(3,3) \  \mathbf{\Delta}_{12}(3,3) 
%\end{gather*}

Now in order to write the parametric form of these matrix elements we demand that all $\Lambda_{\mathbf{\Delta}}$s are positive maps. We would address the complete positivity of these dynamical maps later on. Positivity of the maps can be  written in terms of the following constraints on the elements of their corresponding matrices as:    $\sum_{j = 1}^{3}  \mathbf{\Delta}^{2}(i,j)\leq 1$;  
   $ \sum_{j = 1}^{3} \mathbf{\Delta}^{2}(j,i)\leq 1$, where $\mathbf{\Delta}= \{  \mathbf{\Delta}_{12},\mathbf{\Delta}_{23}  \}$ \cite{Mueller_OSA}. These constraints allow the following parameterization of the matrix elements of the two maps (see Appendix B for details):

\begin{widetext}
    \begin{gather*}
        \begin{matrix}
           \mathbf{\Delta}_{12}(1,3) = r_1 \sin \theta_1 \cos \phi_1 ;&
           \mathbf{\Delta}_{12}(2,3) = r_1 \sin \theta_1 \sin \phi_1; &
           \mathbf{\Delta}_{12}(3,3) = r_1 \cos \theta_1 ;\\
           \mathbf{\Delta}_{23}(3,1) = r_2 \sin \theta_2 \cos \phi_2; &
           \mathbf{\Delta}_{23}(3,2) = r_2 \sin \theta_2 \sin \phi_2; &
           \mathbf{\Delta}_{23}(3,3) = r_2 \cos \theta_2.
        \end{matrix}
    \end{gather*}
\end{widetext}

where $r_1, r_2$ take values between $0$ and $1$, $0 \leq \theta_1, \theta_2 \leq \pi$, $0 \leq \phi_1, \phi_2 \leq 2\pi$.
Using these parameters
% Changed from
% Substituting these values
in the expression of $K_3 = C_{12} + C_{23} - C_{13}$ we find that the LG parameter  takes the following algebraic form:
%\bea
% \begin{equation}
% \begin{split}
%   K_{3} &=& r_1 \cos \theta_1 + r_2 \cos \theta_2 - r_1 r_2 \ [ \cos \theta_1  \cos \theta_2 
%     \nn\\
%         && +   \sin \theta_1  \sin \theta_2 \  \cos (\phi_{1}- \phi_{2})  \ ]
%\end{split} 
%\label{E08}
%\end{equation}
%\eea

\begin{equation}
 \begin{split}
   K_{3} &= r_1 \cos \theta_1 + r_2 \cos \theta_2 - r_1 r_2 \ [ \cos \theta_1  \cos \theta_2  \\
         & +   \sin \theta_1  \sin \theta_2 \  \cos (\phi_{1}- \phi_{2})  \ ]
\end{split} 
\label{E08}
\end{equation}

The maximum possible violation is $3/2$ when parameters $r_1 = r_2 = 1$, $\theta_1 = \theta_2 = \pi/3$ and $|\phi_1 - \phi_2| = \pi$. It should be noticed that for these values of parameters the structure $|C_{ij}|=1/2$  as it was in unitary case. To summarize we have considered two unital maps $\Lambda_{\mathbf{\Delta}_{12}}$ and $\Lambda_{\mathbf{\Delta}_{23}}$, which are only positive (and not necessarily completely positive maps) to calculate the LG parameter. Therefore, the algebraic expression of $K_{3}$ in eqn. (\ref{E08}) defines the bounds for larger set of dynamics than of interest in this paper. Emary in \cite{Emary} demonstrated the violation of LGI for two different scenarios:  (i) where one unital map is assumed to be a unitary map and other map is considered to be non-unitary unital and (ii) two identical non-unitary unital maps are chosen for the dynamics.  On the other hand we have used two generic  unital maps $\Lambda_{\mathbf{\Delta}_{12}}$ and $\Lambda_{\mathbf{\Delta}_{23}}$ which are positive (and will be restricted to completely positive and divisible maps later on) to derive the algebraic expression of LG parameter $K_{3}$.  \\ 

%However, it should be noted that Emary's derivations are based on the fact  that all possible measurements are allowed at different times \cite{Emary}, whereas we have assumed a fixed measurement direction $(\sigma_{z})$ for all times as mentioned earlier. 
We now focus on the nature of two dynamical maps under the constraint that the dynamics lead to maximum violation of LGI.  Using the three matrix elements of $\mathbf{\Delta}_{12}$ with these constraints we can write the action of operator $\mathbf{\Delta}_{12}$ on the $\vec{w}$ written in eqn. (\ref{E3}). Since the outcomes of the measurements along z-direction correspond to $\vec{w}= (0,0,\pm1)$, the resulting states at time $t_{2}$ can be written in terms of $\vec{w}$ as: $\vec{w}(t_{2}) = \pm \ ( \sqrt{3}/2 \cos \phi,\sqrt{3}/2 \sin \phi,1/2 )$ which are pure states. Now this is clear that the states represented by $\vec{w}= (0,0,\pm1)$ remain pure under the evolution defined with the map $\Lambda_{\mathbf{\Delta}_{12}}$. This map $\Lambda_{\mathbf{\Delta}_{12}}$ therefore we call ``unitary like'' for the eigenstates of measurement operator $\sigma_{z}$  only. Next the action of map $\Lambda_{\mathbf{\Delta}_{23}}$ on the states $\vec{w}= (0,0,\pm1)$ can be written as: $\vec{w}(t_{3}) = \pm \ ( \mathbf{\Delta}_{23}(1,3),\mathbf{\Delta}_{23}(2,3),1/2 )$. It is clear from the expression of $\vec{w}(t_{3})$ that the states may not remain pure at time $t_{3}$ immediately before measurement. Therefore, we call $\Lambda_{\mathbf{\Delta}_{23}}$ map a ``non-unitary unital'' map. Recall that positivity of the map demands $\sum_{j = 1}^{3} \mathbf{\Delta}_{23}^{2}(i,j)\leq 1$;  
   $ \sum_{j = 1}^{3} \mathbf{\Delta}_{23}^{2}(j,i)  \leq 1$, hence each element $0 \leq \mathbf{\Delta}_{23}(i,j) \leq 1 $. Using this we can find the tolerance to decohrence of the LG parameter to have maximum value of $3/2$. For this we set $\mathbf{\Delta}_{23}(1,3)=\mathbf{\Delta}_{23}(2,3)=0$ implying
   $\vec{w}(t_{3})= (0,0,\pm1/2)$. Since $|\vec{w}|$ is the length of the Bloch vector, we conclude that $|\vec{w}|\geq 1/2$ defines the tolerance to decohrence for the time interval $t_{3}-t_{2}$. To conclude, this allowed decohrence has to occur in the time interval $t_{3}-t_{2}$ only. Any decoherence in the interval $t_{2}-t_{1}$ fails to violate the LGI upto $3/2$. In the next sub-section we demonstrate this impossibility of obtaining maximum value of $3/2$ for infinitesimally divisible CPTP unital maps. \\

\subsection{Temporal sequencing of unital maps}
\begin{figure}[htb!]
\centering
\includegraphics[width=5.5cm, height=6cm]{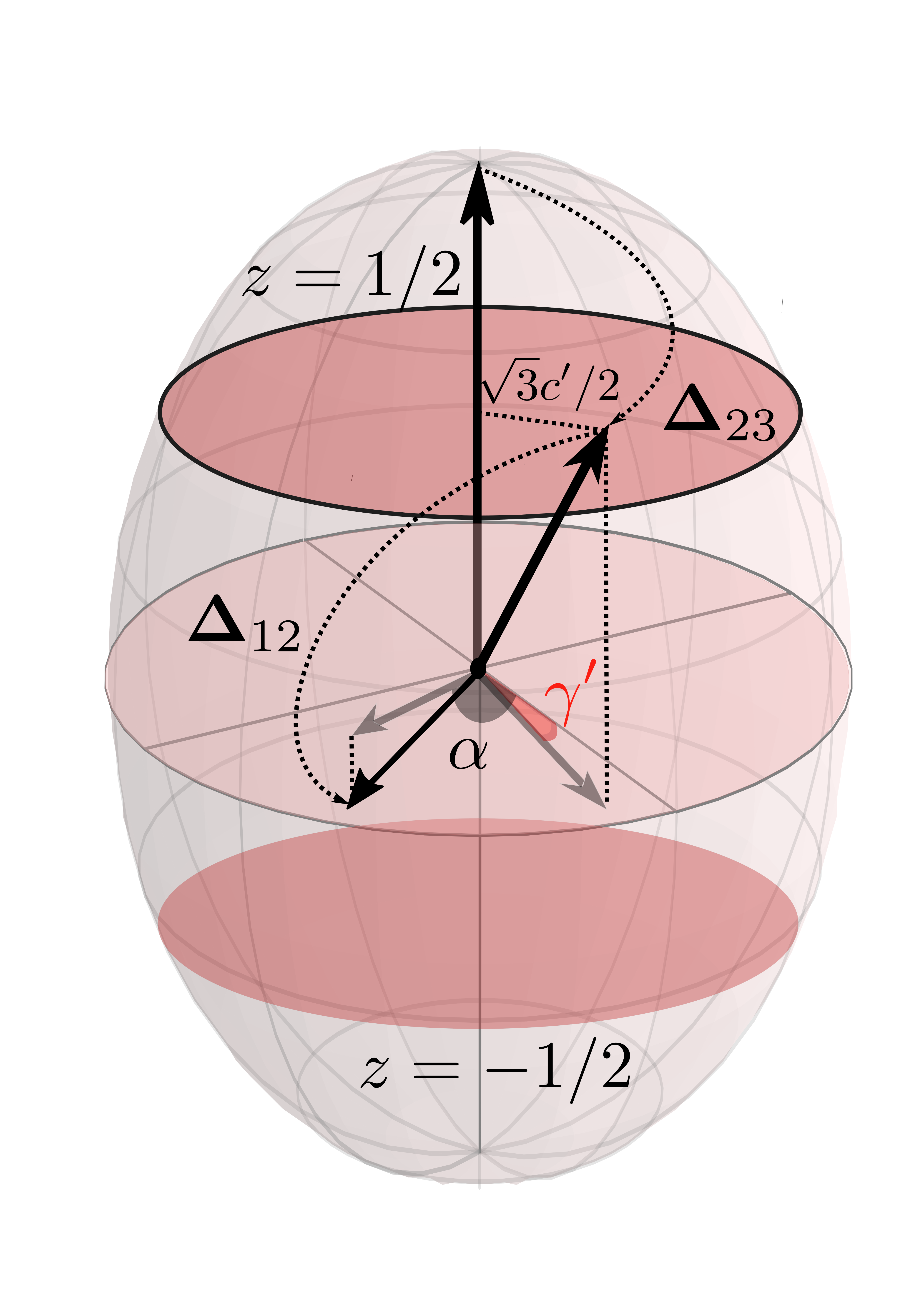}
	\caption{State evolution of eigenstate $| + \rangle_z$ of the measurement operator $\sigma_{z}$ with dynamical map $\Lambda_{\mathbf{\Delta}_{23}}$ and thereafter with $\Lambda_{\mathbf{\Delta}_{12}}$. Grey arrows are in the x-y ($z=0$) geodesic plane. After evolution from time $t_{1}$ to $t_{3}$ with the dynamical maps reversed in order, the Bloch vector has  z-component as $\frac{1}{4} (\ 1-3 \ c  \ c' \cos (\gamma +\text{$\gamma' $}) \ )$, which can never be equal to $-1/2$ as value $c'=1$ is forbidden.}
\label{f2}
\end{figure}

In this section we explicitly write the restricted forms of the two maps $\Lambda_{\mathbf{\Delta}_{12}}$ and $\Lambda_{\mathbf{\Delta}_{23}}$ such that these maps are CPTP divisible, and also allow an extreme violation of LGI upto $K_{3}^{L\ddot{u}ders}$. The above parametrization of the maps is based on the assumption that measurement has been done along z-direction, however the results hold true for any arbitrary direction of measurement as long as it is kept fixed for all times. We provide a geometric visualization of the  impossibility of obtaining $K_{3} = K_{3}^{L\ddot{u}ders}$ when the temporal sequencing of the two maps $\Lambda_{\mathbf{\Delta}_{12}}$ \& $\Lambda_{\mathbf{\Delta}_{23}}$ is reversed (FIG. \ref{f2}). For this purpose,  we decompose the matrices corresponding to these maps in terms of rotation and scaling given by the form: $\mathbf{\Delta} = R_1^T D R_2$ (as mentioned in the last section). We now assume $R_{1}$ and $R_{2}$ are orthogonal matrices representing  rotations and $D$ represents scaling of the Bloch vector $\vec{w}$ i.e. $\text{det}(R_{1})=\text{det}(R_{2})=1$.  Since the map $\Lambda_{\mathbf{\Delta}_{12}}$ corresponding to the matrix $\mathbf{\Delta}_{12}= R_{1}^{T} D_{12} R_{2}$ is ``unitary like'' and hence scaling matrix $D_{12}$ should at least one element that is 1 (see Appendix B for details). Also, complete positivity of this map demands that rest of the two elements should be equal $i.e.$ $D_{12}= \text{diag}(c,c,1)$ and $0 \leq c < 1$. Recall the map is CPTP infinitesimally divisible if $\text{det}(\mathbf{\Delta}) \geq 0$, since $R_{1}$, $R_{2}$ are rotations and $\text{det}(D_{12})>0$, the map $\Lambda_{\mathbf{\Delta}_{12}}$ with these restrictions is not only CPTP but infinitesimally divisible now. Using the fact that  the $D_{12}$ commutes with all rotations about z-axis, $\mathbf{\Delta}_{12}$ can be written as given in eqn. (\ref{E06a}). Using similar argument we can write $\mathbf{\Delta}_{23}$ as in eqn. (\ref{E06a}) (see Appendix B for details), \\

\begin{figure}[htb!]
\centering
\includegraphics[width=5.5cm, height=6cm]{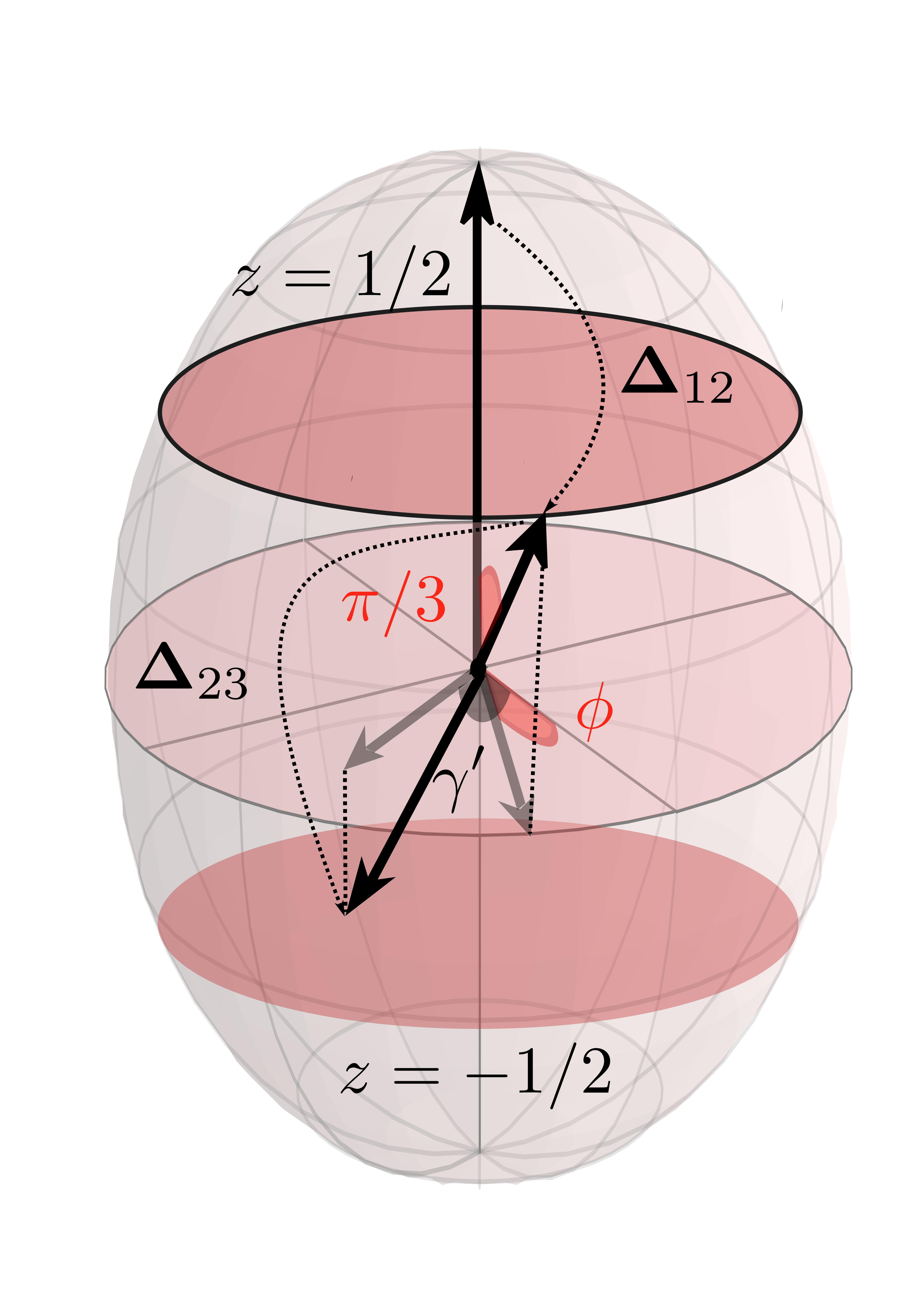}
	\caption{State evolution of eigenstate $| + \rangle_z$ of the measurement operator $\sigma_{z}$ with dynamical map $\Lambda_{\mathbf{\Delta}_{12}}$ and thereafter with $\Lambda_{\mathbf{\Delta}_{23}}$.  Grey arrows are in the x-y ($z=0$) geodesic plane. After evolution from time $t_{1}$ to $t_{3}$ the state $| + \rangle_z$  lies in the $z=-1/2$ plane with Bloch vector $\vec{w}(t_{3}) = \pm \ ( \mathbf{\Delta}_{23}(1,3),\mathbf{\Delta}_{23}(2,3),-1/2 )$ and therefore the maximum of $3/2$ is achieved. }
\label{f1}
\end{figure}

\begin{widetext}

        \begin{gather}
        \label{E06a}
            \mathbf{\Delta}_{12} = R_z (\phi) R_y (\pi / 3) R_z (\gamma) \begin{pmatrix}
                c & 0 & 0 \\
                0 & c & 0 \\
                0 & 0 & 1
            \end{pmatrix} ;  \ 
            \mathbf{\Delta}_{23} = \begin{pmatrix}
                c' & 0 & 0 \\
                0 & c' & 0 \\
                0 & 0 & 1
            \end{pmatrix} R_z (\gamma ') R_y (\pi / 3) R_z (- \phi)
        \end{gather}

\end{widetext}

with arbitrary angles $\gamma, \gamma', \phi$ and $0 \leq c, c' < 1$. The two maps $\Lambda_{\mathbf{\Delta}_{12}}$ and $\Lambda_{\mathbf{\Delta}_{13}}$ defined by matrices $\mathbf{\Delta}_{12}$  and $\mathbf{\Delta}_{23}$ respectively exhaust the full set of CPTP unital infinitesimally divisible maps, which violate the LGI upto $3/2$. In summery, these maps corresponds to Markovian baths for the TLS which allow maximum quantum correlations when used in a correct temporal sequence. Finally, it is important to note that  reversing the temporal sequence of the maps $\Lambda_{\mathbf{\Delta}_{12}}$ and $\Lambda_{\mathbf{\Delta}_{23}}$ and following eqn. (\ref{E05}), we can re-write the correlations $ C_{12}= \mathbf{\Delta}_{23}(3,3) ; \ C_{23} = \mathbf{\Delta}_{12}(3,3)$ and $C_{13}= ( \ \mathbf{\Delta}_{12} \mathbf{\Delta}_{23} \ ) (3,3).$ Recall that for maximum violation correlations should follow the condition $|C_{ij}|=1/2$. And by setting $\mathbf{\Delta}_{23}(3,3)=1/2$ and $\mathbf{\Delta}_{12}(3,3)=1/2$ we can obtain the values of correlations $ C_{12}= 1/2 ; \ C_{23} = 1/2$, which is within the allowed space of parameters constrained by the condition of CPTP and divisible maps. However, obtaining the value $C_{13}=-1/2$ is not possible in this case. In what follows, we demonstrate the possibility of maximum violation of LGI for the temporal sequence $\Lambda_{\mathbf{\Delta}_{23}} \Lambda_{\mathbf{\Delta}_{12}}$ (FIG. \ref{f1}) and impossibility of obtaining $3/2$ when the temporal sequence is reversed (FIG. \ref{f2}), using evolution of measurement operator eigenstate $| + \rangle_z$ on the Bloch sphere for the time interval $t_{1}$ to $t_{3}$ only as we are left with the optimization of the correlation  $C_{13}$. Note that we can fix any two of the three correlations equal to $1/2$, within the constraint of CPTP divisible maps and then focus on the third correlation for the optimization.

As established in eqn. (\ref{E08}) that there is no initial state dependence in the LG parameter, hence we can understand the maximum violation of $K_{3}$ in terms of time evolution of measurement operator eigenstates only. Also, as both the maps are unital we can focus on the evolution of the $| + \rangle_z$ ($\vec{w}(0)= (0,0,+1)$) state for the time interval  $t_{3}-t_{1}$  as the other eigenstate  $| - \rangle_z$ is constrained to evolve as its parity partner $i.e.$ the resulting Bloch vector starting from  $| - \rangle_z$ would be negative of the resulting  vector starting from $| + \rangle_z$ under the action of same unital map. The state $| + \rangle_z$ evolves to a pure state upon application of dynamical map $\Lambda_{\mathbf{\Delta}_{12}}$: length of the Bloch vector remains unity and the resulting state at time $t_{2}$ is $\vec{w}(t_{2}) =  \ ( \sqrt{3}/2 \cos \phi,\sqrt{3}/2 \sin \phi,1/2 )$.
Subsequent evolution of $\vec{w}(t_{2})$ with the dynamical map $\Lambda_{\mathbf{\Delta}_{23}}$ results in the Bloch vector $\vec{w}(t_{3})$ such that z-component is $-1/2$ at time $t_{3}$ which implies that tip of the Bloch vector has to lie in $z=-1/2$ plane as shown in the FIG. (\ref{f1}). This ensures that the correlation $C_{13}$ is pinned to $-1/2$, hence resulting in $K_{3}^{L\ddot{u}ders}$ irrespective of the details of the other parameters.

  However, when the state $| + \rangle_z$ is evolved from time $t_{1}$ to $t_{3}$ now with the temporal sequencing of the unital maps being reversed LG parameter fails to attain maximum value of $3/2$ for any values of parameters in the parameter space define in the eqn. (\ref{E06a}) as shown in FIG. (\ref{f2}). When dynamical map $\Lambda_{\mathbf{\Delta}_{23}}$ is applied, the state $| + \rangle_z$ evolves to a state represented by the Bloch vector $ ( \sqrt{3} \text{c}'/2 \cos (\text{$\gamma $}), \sqrt{3}\text{c}'/2  \sin (\text{$\gamma' $}),1/2)$, where  $0 \leq c' < 1$, which implies the state remains a mixed state as expected.
  Further evolution via dynamical map $\Lambda_{\mathbf{\Delta}_{!2}}$ results in Bloch vector whose  z-component is $\frac{1}{4} (\ 1-3 \ c  \ c' \cos (\gamma +\text{$\gamma' $}) \ )$. Recall that for LG parameter to be maximum the z-component of the Bloch vector must be equal to $-1/2$. Since $c' \neq 1$,  it is impossible for the z-component to be equal to $-1/2$. Hence, in this case LG parameter $K_{3} < K_{3}^{L\ddot{u}ders}$ irrespective of the details of the other parameters.

\section{\label{discussion} Conclusion} 
In conclusion, we identify the full parameter space of dynamical maps which violate the LGI upto L\"{u}ders bound within the constraint that maps  are CPTP divisible (describe Markovian dynamics) and unital in nature. We first find algebraic expression for LG parameter $K_{3}$ for two generic unital positive maps $\Lambda_{\mathbf{\Delta}_{12}}$ and $\Lambda_{\mathbf{\Delta}_{23}}$. Then we constraint the parameter space of the dynamical maps such that $K_{3}^{L\ddot{u}ders}$.
Furthermore, we find that there is a tolerance to decohrence beyond which it is impossible to obtain the L\"{u}ders bound of $3/2$. We also show that presence of decoherence is only allowed in the second evolution corresponding to time interval $t_{3}-t_{2}$, and not in the interval $t_{2}-t_{1}$. We then identify the sub-set of the aforementioned unital dynamical maps, which is CPTP and infinitesimally divisible and violates LGI to its maximum value of $3/2$. Finally, we illustrate the fact that the condition for obtaining $K_{3}^{L\ddot{u}ders}$ is sensitive to the temporal sequence of  maps $i.e.$ decohrence has to occur in the time interval $t_{3}-t_{2}$ only and presence of decoherence in any amount in the interval $t_{2}-t_{1}$ results in failure to attain $K_{3}^{L\ddot{u}ders}$ as depicted in FIG. \ref{f2}.
It is to be noted that Emary has also discussed the case of $K=3/2$  but for restricted situations, which are: (a) evolution for the time interval $t_{2}-t_{1}$ is considered unital and for the time interval $t_{3}-t_{2}$ evolution is taken to be unitary (b) the evolution for both the time intervals namely $t_{2}-t_{1}$ and $t_{3}-t_{2}$ are considered to be identical and ``non-unitary unital'', hence a concrete conclusion regrading temporal sequencing was difficult to establish.

\section{Acknowledgement} 
AVV would like to thank the Council of Scientific and Industrial Research (CSIR), Govt. of India for financial support. S.D. would like to acknowledge the MATRICS grant (MTR/ 2019/001 043) from the Science and Engineering Research Board (SERB) for funding. 

\appendix

\section{Maximal violation of LGI in unitary dynamics}
We choose $\sigma_z$ as our observable. The dichotomic observable $Q_i$ is
defined to be $+1$ if the outcome of the projective measurement is $| \uparrow_z \rangle$ at time $t_i$ and $-1$ otherwise. We choose three
time instants $t_1 \leq t_2 \leq t_3$ for measurements.

A time dependent hermitian dynamics always generates unitary evolution. The state of a system at times $t_i$ and $t_f$ ($t_i \geq t_f$) are related by

\begin{equation}
        | \psi (t_f) \rangle = {\hat{U}}_{t_f, t_i} | \psi (t_i) \rangle
\end{equation}

where $\displaystyle{{\hat{U}}_{t_f, t_i} = \mathcal{T} \exp \left( \frac{1}{i \hbar} \int_{t_i}^{t_f} dt' \hat{H} (t') \right)}$

Any unitary operator for a two level system can be decomposed as \cite{Sakurai-QM}

\begin{equation}
    \hat{U} = \exp{\left( -i \hat{\sigma}_z \phi \right)} \exp{\left( -i \hat{\sigma}_y \theta \right)} \exp{\left( -i \hat{\sigma}_z \xi \right)}
\end{equation}

where $\phi, \theta$ and $\xi$ are real numbers. We will consider that the unitary operator for evolution from $t_{k}$ to $t_{k+1}$ is ${\hat{U}_k = \exp{\left( -i \hat{\sigma}_z \phi_k \right)} \exp{\left( -i \hat{\sigma}_y \theta_k \right)} \exp{\left( -i \hat{\sigma}_z \xi_k \right)}}$ for time instances $t_1, t_2$ and $t_3$.

Suppose that the two level system (we will consider it as a spin-1/2 system) is initially (at time $t_0$) in a state $| \psi (t_0) \rangle$, and it evolves to a state $| \psi (t_1) \rangle$ by some arbitrary Hamiltonian such that if spin is measured along $z$ axis, the probability that it is found to be in $+z$ direction (we will call this event $E_1$) is $\alpha$, and probability to find it to be in $-z$ direction (we will call this event $E_2$) is $\beta$ with $\alpha + \beta = 1$.

%The probabilities of measurement of spin in $z$ direction are $|\alpha|^2$ (when the spin is found to be along $+z$, we will call this event $E_1$) and $|\beta|^2$ (when the spin is found to be along $-z$, which we will call $E_2$).
After the measurement at $t_1$, the state collapses to either $| + \rangle$ (with probability $\alpha$) or $| - \rangle$ (with probability $\beta$) 

If $E_1$ occurs at $t_1$, then probability that $E_1$ occurs at $t_2$ is $\left| \langle + |_z \hat{U}_1 | + \rangle_z \right|^2 = \cos^2 \theta_1$, and probability that $E_2$ occurs at $t_2$ is $\left| \langle - |_z \hat{U}_1 | + \rangle_z \right|^2 = 1 - \left| \langle + |_z \hat{U}_1 | + \rangle_z \right|^2 = \sin^2 \theta_1$.

Given that $E_2$ occurs at time $t_1$, probabilities that $E_1$ occur is
$\left| \langle + |_z \hat{U}_1 | - \rangle_z \right|^2 =  \left| \langle - |_z \hat{U}_1 | + \rangle_z \right|^2 =  \sin^2 \theta_1$. The probability that $E_2$ will occur at $t_2$ is $\left| \langle - |_z \hat{U}_1 | - \rangle_z \right|^2 = \left| \langle + |_z \hat{U}_1 | + \rangle_z \right|^2 = \cos^2 \theta_1$ respectively. Hence, the joint probabilities are
\begin{align*}
    P(+, +) = \alpha \cos^2 \theta_1 \\
    P(+, -) = \alpha \sin^2 \theta_1 \\
    P(-, +) = \beta \sin^2 \theta_1 \\
    P(-, -) = \beta \cos^2 \theta_1 \\
\end{align*}
Hence,
\begin{widetext}

\begin{equation}
    C_{12} = P(+, +) - P(+, -) - P(-, +) + P(-, -) = 2 \left| \langle + |_z \hat{U}_1 | + \rangle_z \right|^2 -1 = \cos 2 \theta_1
\end{equation}

Similarly, $C_{23} = \cos 2 \theta_2$. Note that

\begin{equation}
    \langle + |_z \hat{U}_2 \hat{U}_1 | + \rangle_z = \cos \theta_1 \cos \theta_2 e^{-i \gamma} - \sin \theta_1 \sin \theta_2 e^{i \gamma}
\end{equation}

with $\gamma = \xi_1 + \phi_2$. Hence,

    \begin{equation}
        C_{13} = 2\left| \langle + |_z \hat{U}_2 \hat{U}_1 | + \rangle_z \right|^2 - 1 = \cos 2 \theta_1 \cos 2\theta_2 + \sin 2\theta_1 \sin 2\theta_2 \cos 2 \gamma
    \end{equation}
    \begin{equation}
        K_3 = \cos 2\theta_1 + \cos 2\theta_2 - \cos 2 \theta_1 \cos 2\theta_2 - \sin 2\theta_1 \sin 2\theta_2 \cos 2 \gamma
        \label{K3_herm}
   \end{equation}
\end{widetext}

From eqn. (\ref{K3_herm}), it is evident that the maximum allowed value of $K_3$ is $3/2$ when $\theta_1 = \theta_2 = \pi / 6$ and $\cos 2 \gamma = -1$. Using these values we can see that $C_{12} = C_{23} = -C_{13} = 1/2$.

\section{Maximal violation of LGI in unital dynamics}
Now consider a more general scenario where the evolution is a positive, infinitesimally divisible and linear map. We start with some arbitrary initial density state of the system at time $t = t_0$ and some
arbitrary evolution of the system from $t_0$ to $t_1$. The evolution
of the system for time interval $(t_1, t_2)$ and $(t_2, t_3)$ are considered to be the unital maps given by $3 \times 3$ matrices $\Delta_{12}$ and $\Delta_{23}$ respectively.
Suppose that the system's density state at time $t_1$ in terms of Stokes vector is $(s_1, s_2, s_3)^T$. Then, after measurement it will be projected to either $(0, 0, 1)^T$ with probability $\dfrac{1 + s_3}{2}$ or to the state $(0, 0, -1)^T$ with probability $\dfrac{1 - s_3}{2}$.

The states $(0, 0, \pm 1)^T$ transform to $(\pm \Delta_{12} (1,3), \pm \Delta_{12} (2,3), \pm \Delta_{12} (3,3))^T$ at time $t_2$. The probabilities of measurement outcomes are again determined by the last component of the Stokes vectors. If joint probabilities
of measurement outcomes $(i, j)$ (at $t_1$ and $t_2$ respectively) is denoted as $P(i, j)$ then
\begin{gather*}
    P(+1, +1) = \frac{1 + s_3}{2} \frac{1 + \Delta_{12} (3,3)}{2} \\
    P(+1, -1) = \frac{1 + s_3}{2} \frac{1 - \Delta_{12} (3,3)}{2} \\
    P(-1, +1) = \frac{1 - s_3}{2} \frac{1 - \Delta_{12} (3,3)}{2} \\
    P(-1, -1) = \frac{1 - s_3}{2} \frac{1 + \Delta_{12} (3,3)}{2}
\end{gather*}

The two time correlation function $C_{12}$ is found to be
\begin{widetext}
    \begin{align}
        C_{12} = P(+1, +1) - P(+1, -1) - P(-1, +1) + P(+1, +1) = \Delta_{12} (3,3)
    \end{align}
%\end{widetext}
To be noted that $C_{12}$ does not depend on the initial state at
time $t_0$ or the dynamical map from $t_0$ to $t_1$, which is a
feature of the unital nature of the dynamical maps. Using these results one can find the other two correlation functions

%\begin{widetext}
\begin{gather*}
    C_{23} = \Delta_{23} (3,3) \\
    C_{13} = ( \Delta_{23} \Delta_{12} )_{33} = \Delta_{23} (3,1) \Delta_{12} (1,3) + \Delta_{23} (3,2) \Delta_{12} (2,3) + \Delta_{23} (3,3) \Delta_{12} (3,3)
\end{gather*}

\end{widetext}

Any valid unital map $\Delta$ must map a valid Stoke's vector ($|v| \leq 1$) $v$ to another valid Stokes vector. To ensure this, we must have \cite{Mueller_OSA}
\begin{align*}
    v^T \Delta^T \Delta v \leq 1
\end{align*}
whenever $|v| = 1$. Choosing $v = (1 \ 0 \ 0)^T$, $(0 \ 1 \ 0)^T$ and $(0 \ 0 \ 1)^T$, we get the following set of constraints

\begin{align*}
    \begin{matrix}
    \sum_{j = 1}^{3}  \mathbf{\Delta}^{2}(j,i) \leq 1 & ; &
    \sum_{j = 1}^{3}  \mathbf{\Delta}^{2}(i,j) \leq 1
    \end{matrix}
\end{align*}

The second constraint is derived using $\Delta^T$ instead of $\Delta$, which is also a valid unital map. We write the relevant matrix elements of $\Delta_{12}$ and $\Delta_{23}$ as
\begin{widetext}

    \begin{gather*}
        \begin{matrix}
           \mathbf{\Delta}_{12}(1,3) = r_1 \sin \theta_1 \cos \phi_1 ;&
           \mathbf{\Delta}_{12}(2,3) = r_1 \sin \theta_1 \sin \phi_1; &
           \mathbf{\Delta}_{12}(3,3) = r_1 \cos \theta_1 ;\\
           \mathbf{\Delta}_{23}(3,1) = r_2 \sin \theta_2 \cos \phi_2; &
           \mathbf{\Delta}_{23}(3,2) = r_2 \sin \theta_2 \sin \phi_2; &
           \mathbf{\Delta}_{23}(3,3) = r_2 \cos \theta_2.
        \end{matrix}
    \end{gather*}
\end{widetext}

where $r_1, r_2$ are between $0$ and $1$, $0 \leq \theta_1, \theta_2 \leq \pi$, $0 \leq \phi_1, \phi_2 \leq 2\pi$. Substituting these values in the expression of $K_3 = C_{12} + C_{23} - C_{13}$ we find that the maximum violation possible is $3/2$ when $r_1 = r_2 = 1$, $\theta_1 = \theta_2 = \pi/3$ and $|\phi_1 - \phi_2| = \pi$. With these conditions, the relevant matrix elements become
%\begin{widetext}
%    \begin{gather*}
%        \begin{matrix}
%            \Delta_{12} (1,3) = \dfrac{\sqrt{3}}{2} \cos \phi &
%            \Delta_{12} (1,3) = \dfrac{\sqrt{3}}{2} \sin \phi &
%            \Delta_{12} (3,3) = \dfrac{1}{2} \\
%            \Delta_{23} (3,1) = - \dfrac{\sqrt{3}}{2} \cos \phi &
%            \Delta_{23} (3,2) = - \dfrac{\sqrt{3}}{2} \sin \phi &
%            \Delta_{23} (3,3) = \dfrac{1}{2}
%        \end{matrix}
%    \end{gather*}
%\end{widetext}
\begin{gather}
    \label{mat_elms}
    \begin{matrix}
       \Delta_{12} (1,3) = \dfrac{\sqrt{3}}{2} \cos \phi & \ &
       \Delta_{23} (3,1) = - \dfrac{\sqrt{3}}{2} \cos \phi \\
       \Delta_{12} (2,3) = \dfrac{\sqrt{3}}{2} \sin \phi & \ &
       \Delta_{23} (3,2) = - \dfrac{\sqrt{3}}{2} \sin \phi \\
       \Delta_{12} (3,3) = \dfrac{1}{2} & \ &
       \Delta_{23} (3,3) = \dfrac{1}{2}
    \end{matrix}
\end{gather}
Now we examine what CPTP, divisible maps have these properties. If $\Delta_{12}$ is the matrix of a divisible map then it must be decomposable as ${\Delta_{12} = R_1 \mathcal{D} R_2}$, where $R_1, R_2$ are rotation matrices. Let ${v_z = (0, 0, 1)^T}$. Then, the length of $\Delta_{12} v_z$ is $1$ (using the expressions from eqn. (\ref{mat_elms}), implying at least one of the elements of $\mathcal{D}$ is $1$. And demand for complete positivity following the inequalities $ c_3 + c_2 \leq 1 + c_1$  and $1 \geq c_3 \geq c_2 \geq |c_1|$ implies that the other
two elements must be equal. Thus, $\mathcal{D}$ must be of the form ${\text{diag}(c, c, 1)}$ where $c$ is a real number between $0$ and $1$. This allows us to further subdivide these maps into two classes, (a) unitary, when $c = 1$ and (b) non unitary when $0 \leq c < 1$. For unitary dynamics, the general expression for getting maximal violation is already discussed. Since the image of $v_z$ has same length as $v_z$, $R_2 v_z$ must be an eigenvector of $\mathcal{D}$ with eigenvalue $1$. Hence, without loss of generality we can consider $R_2$ to be a rotation about the $z$ axis on the Bloch sphere. We can decompose the other rotation $R_1$ with Euler angles as $R_1 = R_z(\alpha) R_y(\beta) R_z(\gamma)$. Then, $R_1 v_z = (\sin \beta \cos \alpha, \sin \beta \sin \alpha, \cos \beta)$. Hence, without loss of generality we can claim that
$R_1 = R_z(\alpha) R_y(\pi / 3) R_z(\gamma)$ with arbitrary $\gamma$. And since rotation about $z$ axis commutes with $\mathcal{D}$, we can as well choose $R_2$ to be identity.

Similarly, we observe that the length of $v_z^T \Delta_{23}$ is $1$. And we can parameterize the map with same parameters in reverse order.  Hence, the matrices of all possible CPTP divisible maps that give maximal violation of $K_3$ are of the form
\begin{widetext}
    \begin{subequations}
        \begin{gather}
           \mathbf{\Delta}_{12} = R_z (\phi) R_y (\pi / 3) R_z (\gamma) \begin{pmatrix}
                c & 0 & 0 \\
                0 & c & 0 \\
                0 & 0 & 1
            \end{pmatrix} \\
            \mathbf{\Delta}_{23} = \begin{pmatrix}
                c' & 0 & 0 \\
                0 & c' & 0 \\
                0 & 0 & 1
            \end{pmatrix} R_z (\gamma ') R_y (\pi / 3) R_z (- \phi)
        \end{gather}
    \end{subequations}
\end{widetext}
with arbitrary angles $\gamma, \gamma', \phi$ and $0 \leq c, c' < 1$.

%To be noted that $K$ rotates the eigenstates of the measurement operator $\sigma_z$ on the Bloch sphere coherently. While the same is not true for $M$. Also, for maximal violation of $K_3$, we always have $C_{12} = C_{23} = - C_{13} = 1/2$. Geometrically, both $K$ an $M$ evolves the states $(0, 0, \pm 1)$ on the intersection of the plane $z = \pm 1/2$ and the Bloch ball. While $M$ and $K$ combined evolves $(0, 0, \pm 1)$ on the intersection of Bloch ball and the $z = \mp 1/2$ plane.

For maximal violation, we will always have $C_{12} = C_{23} = - C_{13} = 1/2$. And $\Delta_{12}$ rotates the eigenstates of the measurement operator coherently, which is not necessarily true for both $\Delta_{23}$ and $( \Delta_{23} \Delta_{12})$. This means that the state $| \uparrow_z \rangle$ at $t = t_1$ will be a pure state at the intersection of Bloch sphere and $z = 1/2$ plane at time $t_2$. When the measurement is performed at $t = t_2$ and the outcome is $| \uparrow \rangle_z$, the state will be in the intersection of the Bloch ball and $z = 1/2$ plane at $t = t_3$. When measurement outcome at $t = t_1$ is $| \uparrow_z \rangle$ and no measurement is done at $t = t_2$, the state will be found at the intersection of the Bloch ball and $z = -1/2$ plane.

\end{document}